\documentstyle[preprint,aps]{revtex}
\begin{document}
\preprint{UM-P-97/12, RCHEP-97/3}
\draft
\title{Electron-positron annihilation into Dirac magnetic monopole and
antimonopole: the string ambiguity and the discrete symmetries}
\author{A.Yu.Ignatiev\cite{byline1} and G.C.Joshi\cite{byline2}}
\address{Research Centre for High Energy Physics, School of
Physics, University of Melbourne, Parkville, 3052, Victoria,
Australia}
\maketitle
\begin{abstract}
We address the problem of string arbitrariness in the quantum field
theory of Dirac magnetic monopoles. Different prescriptions are shown
to yield different physical results. The constraints due to the
discrete symmetries (C and P) are derived for the process of electron-
positron annihilation into the monopole-antimonopole pair.
In the case of the annihilation through the one-photon channel, the
production of spin 0 monopoles is absolutely forbidden; spin 1/2
monopole and antimonopole should have the same helicities (or,
equivalently, the monopole-antimonopole state should be p-wave
$^1P_1$).
\end{abstract}
\pacs{14.80.Hv}

1.The experimental searches for Dirac magnetic monopoles produced (as
real or virtual particles) in $e^+e^-$ annihilation have been
conducted for several decades. For reviews on Dirac monopoles \cite{D}
see, e.g., \cite{ST,c,rev}; recently, there have been discussions of
the effects of virtual monopoles    on the anomalous magnetic moment
of the electron \cite{amm} and in Z-boson decay \cite{r}.
If we are to derive an upper bound on the monopole mass based on the
negative results of these searches, we have to be able to calculate
theoretically the amplitude of the monopole-antimonopole production in
$e^+e^-$ annihilation for a given monopole mass. However, this
calculation presents serious problems.

There are two sources of difficulties. The first
is a well-known fact that the coupling constanst (that is, the
magnetic charge of the pole) should be very large if the Dirac
quantization condition is to be true. That makes impossible the use of
perturbation theory for practical calculations (although it
can be used within an effective field theory approach).
The second difficulty
(which has not been as much popularized) is, perhaps, more
fundamental. It has nothing to do with the magnitude of the coupling
constant at all; rather, it is related to the existence of Dirac
string -- infinitely thin line of magnetic flux stretching from the
magnetic pole to infinity. It has been often repeated in the
literature that the Dirac quantization condition makes the string
invisible. However, in reality the situation is far from being so
simple and clear. This is especially true in the context of quantum
field theory where monopoles are allowed to be created and
annihilated (recall that the Dirac quantization condition was                  
initially derived for a simple {\em quantum mechanical} system ``
electron plus monopole''). It is generally believed that the full
quantum field theory does not depend on how we choose the position of
the string which can be arbitrary. However, the peculiarity of the
monopole theory is that the formulation of the theory cannot be made
without recourse to the string concept in one or another form. In
other words, the quantum theory of monopoles is not {\em manifestly}
string-independent. Since the string fixes a specific direction, the
theory is not {\em manifestly} Lorentz invariant either. Perhaps, it
is a unique example of a physical theory possessing implicit Lorentz
invariance which nevertheless cannot be formulated in a {\em
manifestly} invariant way.

What are the practical implications of this fundamental theoretical
feature? One consequence is this. Imagine that we forget for a moment
about the large coupling constant and attempt to calculate some
physical quantity in the first order of perturbation theory (such as
the monopole-antimonopole production in $e^+ e^- $ annihilation). The
result will be discouraging because it will be ambiguous. More
exactly, the result will depend explicitely on the string direction
which is clearly unacceptable. Obviously, a serious question is how to
deal with this type of situation.
Consider, for example, the process of $e^+e^-$ annihilation
into monopole-antimonopole pair (assumed to be fermions). It has a
virtue of being physically interesting and simple enough at the same
time. This process has been previously considered and a prescription
has been given for elimination of string dependence \cite{d} which
has
been subsequently adopted in \cite{amm,r}. The resulting
cross-section is not very
different from the cross-section for the creation of a pair of usual
fermion-antifermion.

However, we believe that the prescription is not enirely
satisfactory. One reason for concern is that it only
gives the value of
the {\em squared modulus of the amplitude}, but not {\em the
 amplitude} itself.
Therefore, it would be difficult to generalize it for the cases
when an {\em interference} of two amplitudes is involved (for
instance, if we want to calculate the interference
between electromagnetic and Z-boson contributions to the
monopole-antimonopole production in $e^+e^-$ annihilation).

One purpose of this paper is to consider an alternative
 procedure and see if the physical results would be the same.  More
specifically, we  propose an alternative prescription
based on the averaging of the amplitude over all possible directions
of the
string. This procedure has a clear physical meaning since the string
is supposed to be unobservable. However, it leads to a drastically
different answer: according to this prescription, the amplitude
of $e^+e^-$ annihilation into the monopole-antimonopole pair
should be {\em zero} to the lowest order of perturbation theory.

This result suggests that the task of extracting the physically
meaningful results from the inherently ambiguous perturbative
calculations should be considered as an open problem requiring
further investigation.                                 
In this paper we try to circumvent this problem by using only general
 principles
of quantum field theory whose validity does not rely on the use
of perturbation theory. It is natural to start with the consideration
of the role of the discrete symmetries such as C and P
transformations and to see what constraints are provided
by these symmetries.

We show that the behaviour of the monopole-antimonopole
system under discrete symmetries
is rather different from that of standard fermion-antifermion
or boson-antiboson system (standard means
not carrying magnetic charge). In particular, there arise
selection rules for the process of the monopole-antimonopole
production through one-photon annihilation of an electron
and positron. For spin 1/2 monopole the P and C symmetries
require that the monopole
and antimonopole have {\em the same} helicities.
For {\em spinless} monopoles
CP symmetry {\em absolutely forbids} the monopole-antimonopole
production through the one-photon annihilation
of an electron and positron.

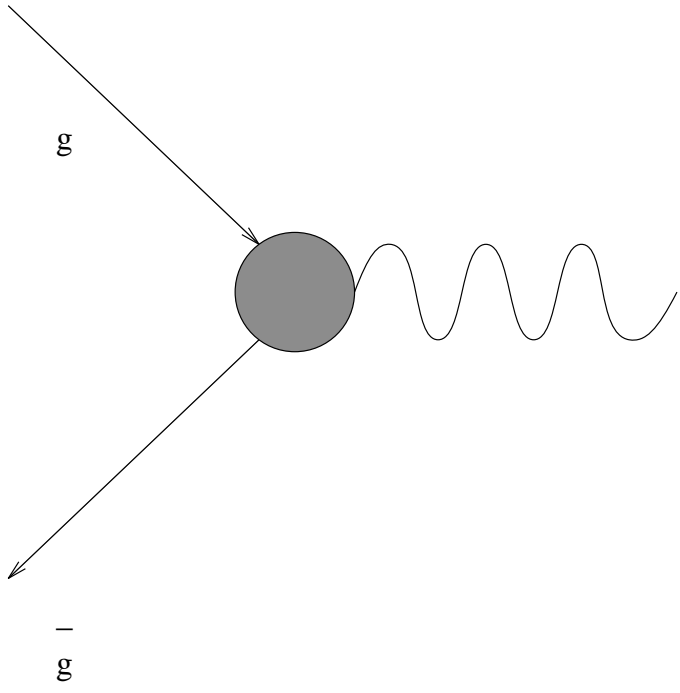
\begin{figure}
\caption{Feynman rules for the photon-monopole interaction}
\label{fig1}
\end{figure}

2.The Feynman rules \cite{d} describing the interactions of photons
and monopoles
have the following form (Fig.~\ref{fig1}):
\begin{equation}
-ig{\epsilon^{\mu\nu\lambda\rho}\gamma^{\nu}n^{\lambda}q^{\rho}
\over qn+i \epsilon}.
\end{equation}
The photon and fermion propagators, as well as the photon-electron
vertex,
remain the same as in the standard QED.
Note that in other formulations of the monopole quantum field theory
the Feynman rules would be different (for details, see \cite{rev}).
The most notable feature of these Feynman rules is the fact that they
depend on the vector $n$ which corresponds to the direction of the
string. Thus, these Feynman rules are not manifestly invariant.
However, it is believed that the full theory is nevertheless Lorentz-
invariant, that is physical predictions should not depend on the
specific direction of the vector $n$.

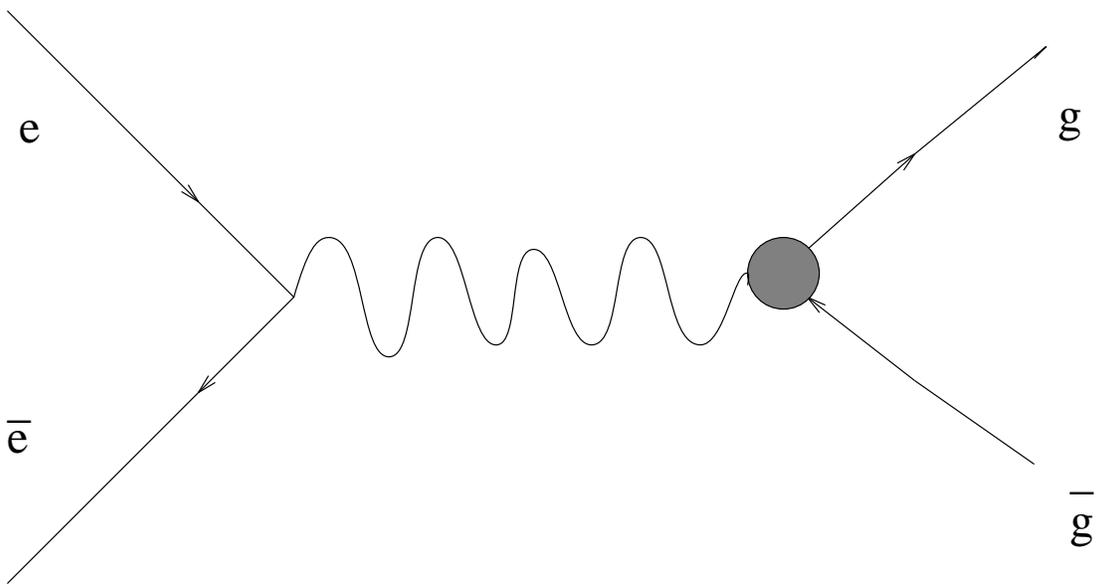
\begin{figure}
\caption{Electron-positron annihilation into monopole-antimonopole}
\label{fig2}
\end{figure}

Now, let us write down the amplitude \cite{d} of the process of the
electron-
positron annihilation into the monopole-antimonopole pair. The
amplitude has the following form (Fig.~\ref{fig2}):
\begin{equation}
A=ieg K^{\beta}
\epsilon^{\mu\beta\gamma\delta}
{n^{\gamma}q^{\delta} \over qn}{1 \over q^2}
J^{\mu}.
\end{equation}
where
\begin{equation}
J^{\mu}={\bar v}_e(p_2)\gamma^{\mu}u_e(p_1),
K^{\beta}={\bar u}_g(p_3)\gamma^{\beta}v_g(p_4).
\end{equation}                               
The dependence on $n$ remains even after the squaring of the amplitude
is made. An obvious question is how to make sense out of the n-
dependent quantity. It has been suggested in Ref.~\cite{d} that one
should
drop the terms which have no pole in $q^2$ and thus to arrive at the
following result:
\begin{equation}
\label{5}
|A|^2={e^2g^2 \over q^4}[(KJ^{\dag})(JK^{\dag})-
(JJ^{\dag})(KK^{\dag})].
\end{equation}

3.However, the consistency of such a prescription can be questioned
on the grounds that it gives the corrected value of the
{\em squared matrix element} but not of the {\em amplitude} itself.
Therefore, it would be difficult to generalize it for the cases
when an {\em interference} of two amplitudes is involved (for
instance, if we want to calculate the interference
between electromagnetic and Z-boson contributions to the
monopole-antimonopole production in $e^+e^-$ annihilation).
Another concern is whether Eq.~ (\ref{5}) is positively
definite or not.
There exists a different approach to the problem of dealing with
the $n$ dependence. The idea is to average over all possible
directions of $n$. Since there are no physically preferred directions
of $n$, all the directions should be taken with the same weight.
Because all these directions are physically indistinguishable, we have
to perform averaging of the amplitude rather than of the squared
 matrix
element. Therefore, we need  to find the average value:
\begin{equation}
\langle {n^{\gamma} \over qn} \rangle.
\end{equation}
By Lorentz invariance, it is sufficient to find this average value
in a system where $n^0=0$ and, consequently, ${\bf n}^2=1$:

\begin{equation}
\langle {{\bf n} \over - {\bf q}{\bf n}}\rangle =
{1 \over 4\pi} \int {{\bf n} \over - {\bf q}{\bf n}}d\Omega.
\end{equation}
In evaluating this integral one should be careful about a possible
singularity arising when the vector ${\bf n}$ becomes orthogonal
to ${\bf q}$. Let us choose the $z$ axis of the spherical coordinate
system such as to be parallel to ${\bf q}$, and calculate the $x,y,z$
components of the average:
\begin{equation}
{1 \over 4\pi}\int {n_x \over - {\bf q}{\bf n}} d\Omega =
-{1 \over 4\pi|{\bf q}|}\int^{1}_{-1} {\sqrt{1-t^2} \over t}dt
\int^{2\pi}_{0}\cos\phi d\phi ,
\end{equation}
where $t=\cos\theta$. Although the integral over $\phi$ vanishes,
we need to prove that the integral over $t$ is not singular.
For this purpose we have to invoke the $ qn+i\epsilon$ rule
(or, in 3-dimensional terms, the $ {\bf q}{\bf n} -i\epsilon$ rule):
\begin{equation}
\int^{1}_{-1} {\sqrt{1-t^2} \over t}dt \rightarrow
\int^{1}_{-1} {\sqrt{1-t^2} \over t -i\epsilon}dt =
\wp \int^{1}_{-1} {\sqrt{1-t^2} \over t}dt +                                
i\pi \int^{1}_{-1} \delta (t) \sqrt{1-t^2}dt =i\pi
\end{equation}
Thus, indeed, the $t$-integral is finite and, therefore,
\begin{equation}
{1 \over 4\pi}\int {n_x \over - {\bf q}{\bf n}} d\Omega =0.
\end{equation}
Furthermore, a similar argument shows that the $y$-component
of the average  value also vanishes:
\begin{equation}
{1 \over 4\pi}\int {n_y \over - {\bf q}{\bf n}} d\Omega =0.
\end{equation}
Now, the $z$-component is;
\begin{equation}
{1 \over 4\pi}\int {n_z \over - {\bf q}{\bf n}} d\Omega =
-{1 \over 4\pi} {1 \over |{\bf q}|} \int d\Omega =
-{1 \over |{\bf q}|}.
\end{equation}
Thus, finally, we obtain:
\begin{equation}
{1 \over 4\pi}\int {{\bf n} \over - {\bf q}{\bf n}} d\Omega =
- {{\bf q} \over {\bf q}^2}.
\end{equation}
Consequently,
\begin{equation}
\langle {n^{\gamma} \over qn} \rangle={q^{\gamma} \over q^2}.
\end{equation}
Now, if we insert this value into the $e^+e^-$ annihilation
amplitude, we obtain
\begin{equation}
A=ieg K^{\beta}
\epsilon^{\mu\beta\gamma\delta}
{q^{\gamma}q^{\delta} \over q^4}
J^{\mu}=0.
\end{equation}
Thus, we arrive  to the conclusion: {\em  if one uses the
averaging procedure to eliminate the string dependence of the
 amplitude,
than  the one-photon
amplitude of the $e^+e^-$ annihilation into monopole-antimonopole pair
turns out to be zero}.
To summarize, we have shown that two different prescriptions used
to eliminate the string dependence of the amplitude lead to
drastically different physical results. Therefore, we suggest
to try to circumvent this problem by using only general principles
of quantum field theory whose validity does not rely on the use
of perturbation theory. It is natural to start with the consideration
of the role of the discrete symmetries such as C, P and T
transformations and to see what constraints are provided
by these symmetries.

4. There exist several formulations of the quantum field theory with
electric and magnetic charges. However, all of the formulations have
been shown \cite{rev}
to be equivalent (except the formulation due to Cabibbo and
Ferrari). Therefore, we will not need to specify exactly in which
theoretical context are going to work. Rather, we will focus on the
properties of the quantum field theory under the action of the
discrete symmetries such as space reflection and charge conjugation.
It can be shown \cite{ram,w,z}                                
that the quantum field theory of the electric and
magnetic charges is invariant under the following discrete
transformations (we use Majorana representation, and denote the
magnetically charged fields by the subscript $g$):
\begin{equation}
C:\;\;{\bf E},{\bf H},\psi,\psi_g \rightarrow -{\bf E},-{\bf H},
\psi^{\dag},\psi_g^{\dag}.
\end{equation}
\begin{equation}
\label{17}
P:\;\;{\bf E(x)},{\bf H(x)},\psi{\bf (x)},\psi_g{\bf (x)} \rightarrow
-{\bf E(-x)},-{\bf H(-x)},\gamma^0\psi{\bf (-x)},\gamma^0
\psi_g^{\dag}{\bf (-x)}
\end{equation}
\begin{equation}
T:\;\;{\bf E}(t),{\bf H}(t),\psi(t),\psi_g(t) \rightarrow {\bf E}(-
t),-{\bf H}(-t),\gamma^0 \gamma^5
\psi (-t),\gamma^0 \gamma^5\psi_g^{\dag}(-t).
\end{equation}
Here, a comment on terminology is in order. There is some confusion in
the literature as to whether we should retain the names ``P
reflection'' and ``T inversion'' for the above operations or we should
call them ``PM'' and ``TM'' transformations, where M stands for the
inversion of the magnetic charge. However, this difference is of
semantical rather than of physical character;  switching from one
terminology to the other does not entail any physical consequences. In
this paper we adopt the the first point of view, i.e. we keep the
names parity and T inversion for the operations we have just
introduced without making any further qualifications (the same view
is adopted in \cite{w}).

Sometimes one can find in the literature the statements to the effect
that the theory of monopoles is not invariant under P and T
symmetries. These statements refer to the situation when the discrete
symmetries are assumed to act on the magnetically charged particles in
exactly the same way as they act on the electrically charged
particles, that is their action on the magnetically charged states
does not include the sign inversion of the magnetic charge. It is easy
to see that if the discrete transformations are defined in that way,
then the theory is indeed P and T non-invariant.  However, the
possibility to define P and T symmetries in such a way that they are
conserved makes the ``non-conserving'' definition irrelevant.

Note also that we assume that there are no particles carrying
simultaneously both the electric and magnetic charge; in other words,
there are no dyons in the theory; in this case the conserving P and T
operations do not exist \cite{z}.

Now, in terms of creation (or annihilation ) operators (rather than in
terms of local fields) the transformation law for a magnetically
charged fermion takes the form:
\begin{equation}
Pa_g(p,s)P^{-1}=b_g(-p,s),\;\;\; Pb_{g}^{\dag}(p,s)P^{-1}=
-a_{g}^{\dag}(-p,s).
\end{equation}
For a magnetically uncharged fermion $\psi$, the transformation law
is:
\begin{equation}
Pa(p,s)P^{-1}=a(-p,s), \;\;\; Pb(p,s)P^{-1}=-b(-p,s).                     
\end{equation}
The law of C transformation of a magnetically charged fermion has the
same form as for a fermion without magnetic charge:
\begin{equation}
Ca_g(p,s)C^{-1}=b_g(p,s),  \;\;\; Cb_{g}(p,s)C^{-1}=
a_{g}(p,s).
\end{equation}
Similarly, one can obtain the formulas for the T reversal but we will
not need to use them in the present paper. Hence, we see a clear
difference between the behavior of the states with the electric charge
and the magnetically charged states. The parity and time inversion
acting on
the electrically charged states do not change the electric charge of
these states, that is under P transformation the electron is carried
into an electron with opposite momentum and, likewise, positron is
transformed into positron state with the opposite momentum. Similarly,
under time inversion the electron state is transformed into the
electron state with opposite momentum and spin; the positron is turned
into the positron with opposite momentum and spin. So, the P and T
transformation do not change the electric charge at all. On the
contrary, for magnetically charged particles the situation is exactly
opposite: the P and T reflections necessarily include {\em the
change of
sign} of the magnetic charge. For instance, P transformation acting on
the magnetic monopole
takes it into antimonopole with the opposite momentum; likewise, under
P parity the antimonopole is transformed into monopole with the
opposite momentum. The same is true for T reversal: the T
transformation changes the monopole into antimonopole with opposite
momentum and spin; the antimonopole is changed into monopole with
inverse momentum and spin.

5. Now we are in a position to apply the discrete symmetries
to consideration of specific physical processes in order
to establish whether any selection rules can be obtained
or not.
Since the monopoles are expected to be relativistic, let us
use the helicity basis for their consideration. In this
basis the pair of monopole-antimonopole is described by
a wave function $\psi_{JM\lambda_1\lambda_2}$ where $J$ is
the total angular momentum of the pair, $M$ is the projection
of $J$ and $\lambda_1$ and   $\lambda_2$  are the helicities of
the monopole and antimonopole.
The action of discrete symmetries is given by:
\begin{equation}
P\psi_{JM\lambda_1\lambda_2}=\psi_{JM-\lambda_2 -\lambda_1},
\end{equation}
\begin{equation}
C\psi_{JM\lambda_1 \lambda_2}=(-1)^J\psi_{JM\lambda_2 \lambda_1}.
\end{equation}
Using these rules, we can construct the wave function that has
the photon quantum numbers, i.e. $J=1$, $P=-1$ and $C=-1$:
\begin{equation}
\label{wf}
\psi_{1M}={1\over \sqrt{2}}(\psi_{1M {1 \over 2}{1 \over 2}}
            -\psi_{1M -{1 \over 2}-{1 \over 2}}).
\end{equation}
Thus we see that in order to couple to the photon, the monopole                
and antimonopole should have the {\em same} helicities.
To further understand the physical meaning of this condition,
let us consider the non-relativistic limit, in which the monopole
-antimonopole pair is described by the wave function
$\psi_{JLSM}$ where $L$ and $S$ are the total orbital momentum
and spin, respectively. The connection between the wave functions
$\psi_{JLSM}$ and $\psi_{JM\lambda_1\lambda_2}$ is given by
\cite{BLP}:
\begin{equation}
\psi_{JLSM}=\sum_{\lambda_1\lambda_2}
\psi_{JM\lambda_1\lambda_2}
\langle JM \lambda_1\lambda_2 | JLSM \rangle,
\end{equation}
where the coefficients are expressed through the $3j$ symbols
as follows:
\begin{equation}
\langle JM \lambda_1\lambda_2 | JLSM \rangle=
(-i)^L(-1)^S \sqrt{(2L+1)(2S+1)}
\left( \begin{array}{ccc}
{1 \over 2}&{1 \over 2}&S \\
\lambda_1& -\lambda_2&- \Lambda
\end{array} \right)
\left( \begin{array}{ccc}
L&S&J\\
0& \Lambda &-\Lambda
\end{array} \right).
\end{equation}

It can be shown that the wave function Eq.~(\ref{wf}) corresponds
to the state $S=0$, $L=1$ in the non-relativistic limit, i.e.:
\begin{equation}
\psi_{110M} = {1\over \sqrt{2}}(\psi_{1M {1 \over 2}{1 \over 2}}
            -\psi_{1M -{1 \over 2}-{1 \over 2}}),
\end{equation}
\begin{equation}
\Lambda=\lambda_1-\lambda_2
\end{equation}
In other words, the $J^{PC}=1^{--}$ of the monopole-antimonopole
pair corresponds to the $^1P_1$ state in the non-relativistic
limit. This should be contrasted with the case of the standard
fermion-antifermion pair (such as positronium or quarkonium)
for which the $1^{--}$ state is $^3S_1$ (or $^3D_1$).

Now, let us consider spin 0 monopoles for we do not have any
evidence concerning the possible value of the monopole spin.
From the similar considerations as the above, it can be shown
that the spinless monopole-antimonopole system has the following
quantum numbers:
\begin{equation}
\label{s}
P=1, \;\;\; C=(-1)^J,
\end{equation}
where $J$ is the total angular momentum of the system.
Thus, the spinless monopole-antimonopole production through the
one-photon $e^{+}e^{-}$ annihilation is {\em absolutely
forbidden}.
Next, it follows from Eq.~(\ref{s})
that in the state with the total angular momentum
$J=1$ the monopole-antimonopole pair has always $CP=-1$.                    
Therefore, CP symmetry absolutely forbids the $1^{--}$
and $1^{++}$ states of the monopole-antimonopole system.
Note that this conclusion holds true even if P and C parities
do not conserve separately, but CP does.
This means that the the decay of Z-boson into  spin 0
 monopole-antimonopole pair would be absolutely forbidden in
a CP invariant theory.

Thus we have shown that C and P invariance imposes exact
selection rules on the monopole-antimonopole state produced
through the one-photon channel of $e^+e^-$ annihilation.

It remains to be investigated whether these selection rules
can help us to understand why the monopoles have not been
observed experimentally.

Recently the contribution of virtual monopoles to various physical
processes has been examined in several papers. One of them was the
contribution of virtual monopole-antimonopole pairs to the anomalous
magnetic moment of the electron \cite{amm} (see Fig.3).
Another process is the monopole loop
contribution to the decay of Z boson into 3 photons \cite{r}.
 By necessity, the calculation of all these diagrams
involves the use of some prescription to eliminate the string
dependence of physical results. As we have seen,  the validity of
the conclusions drawn with the help
of such prescriptions remains uncertain and needs to be further
investigated.

6.To summarize, we have examined several issues related to the
processes of Dirac
monopole-antimonopole production in high-energy collisions
such as $e^+e^-$ annihilation.
Perturbative calculations for such processes are known to be
inherently ambiguous due to the arbitrariness of direction
of the monopole string; this requires use of some prescription
to obtain physical results.
We argue that different prescriptions lead to  drastically
different physical results which suggests that at present we
do not have an entirely satisfactory procedure for the
elimination of string arbitrariness (this problem is quite
separate from the problems caused by the large coupling
constant). We then analyze the consequencies of discrete
symmetries (P and C) for the monopole production processes and for
the monopole-antimonopole states. The P and C selection rules
for the monopole-antimonopole states turn out to be different
from those for the ordinary fermion-antifermion or boson-antiboson
 systems.
In particular, the spin 1/2 monopole and antimonopole should have
{\em the same} helicities if they are produced through
one-photon annihilation of an electron and positron.
In the case of  {\em spinless} monopoles
CP symmetry {\em absolutely forbids} the monopole-antimonopole
production through the one-photon  annihilation
of an electron and positron.
This work was supported in part by the Australian Research Council.

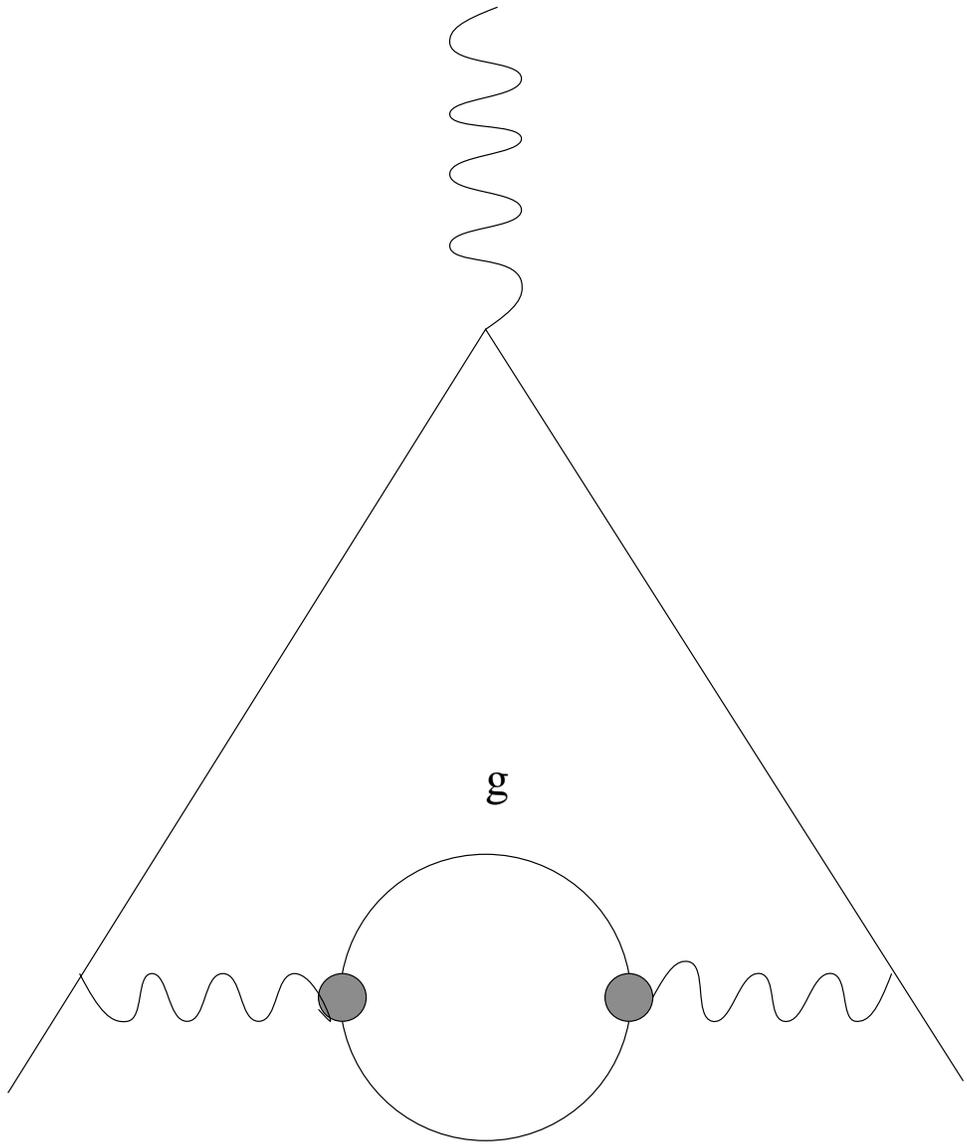
\begin{figure}
\caption{Contribution of virtual monopole-antimonopole pair to the
anomalous
magnetic moment of the electron}
\label{fig5}
\end{figure}

\end{document}